\documentclass{kluwer}    
\usepackage[dvips]{graphicx}

\newdisplay{guess}{Conjecture}

\begin{document}                                                                                   
\begin{article}
\begin{opening}         
\title{RHESSI as a Hard X-Ray Polarimeter} 
\author{Mark L. \surname{McConnell}} 
\author{James M. \surname{Ryan}}  
\institute{Space Science Center, University of New Hampshire, Durham, NH  03824}
\author{David M. \surname{Smith}}  
\institute{Space Sciences Laboratory, University of California, Berkeley, CA  94720}
\author{Robert P. \surname{Lin}}  
\institute{Physics Department and Space Sciences Laboratory, University of California, Berkeley, CA  94720}
\author{A. Gordon \surname{Emslie}}  
\institute{Physics Department, University of Alabama, Huntsvile, AL  35899}

\runningauthor{McConnell et al.}
\runningtitle{RHESSI as a Polarimeter}
\date{September 10, 2002}

\begin{abstract}
Although designed primarily as a hard X-ray imager and spectrometer, the Ramaty High Energy Solar Spectroscopic Imager (RHESSI) is also capable of measuring the polarization of hard X-rays (20--100 keV) from solar flares.  This capability arises from the inclusion of a small unobstructed Be scattering element that is strategically located within the cryostat that houses the array of nine germanium detectors.  The Ge detectors are segmented, with both a front and rear active volume.  Low energy photons (below about 100 keV) can reach a rear segment of a Ge detector only indirectly, by scattering.  Low energy photons from the Sun have a direct path to the Be and have a high probability of Compton scattering into a rear segment of a Ge detector. The azimuthal distribution of these scattered photons carries with it a signature of the linear polarization of the incident flux.  Sensitivity estimates, based on Monte Carlo simulations and in-flight background   measurements, indicate that a 20--100 keV polarization sensitivity of less than a few percent can be achieved for X-class flares.  
\end{abstract}

\end{opening}           

\section{Introduction}
\label{sect:intro}  

Hard X-ray emission from solar flares, like any other form of electromagnetic radiation, has four, and only four properties.  Each photon can be completely characterized by its time of arrival, its energy, its direction of arrival, and its polarization state.  RHESSI is capable, to varying degrees, of determining all four of these quantities. The first two of these properties (time of arrival and energy) are measured directly as photon interactions in the Ge detectors. The third and fourth properties (arrival direction and polarization state) are determined through an analysis of the grouping of photons in time at each detector.  RHESSI is capable of providing information on the linear polarization of photons between roughly 20 and 100 keV.  The study of polarization at hard X-ray energies is especially appealing in that the hard X-ray emission from any bremsstrahlung source (such as a solar flare) will be polarized if the phase-space distribution of the emitting electrons is anisotropic.  Polarization measurements therefore provide a direct handle on the extent to which the accelerated electrons are beamed, which, in turn, has important implications for particle acceleration models.  

Studies of $\gamma$-ray line data from the SMM Gamma Ray Spectrometer (GRS) suggest that protons and $\alpha$-particles are likely being accelerated in a rather broad angular distribution \cite{Share97,Share02}.  There is no reason to expect, however, that electrons are being accelerated in a similar fashion.  The ability of RHESSI to make polarization measurements, coupled with its ability to (independently) image the hard X-ray emission, provides a powerful tool for studying electron acceleration in solar flares.  In this section, we  first review the current status of efforts to measure the electron beaming (via photon directivity) using studies of the distribution of observed events across the solar disk and stereoscopic studies of individual flares.  We then review the potential value of polarization measurements and the checkered history of polarization studies of solar flares. 

\subsection{Hard X-Ray Directivity Measurements}
Related to polarization is the directionality of the hard X-ray radiation yield.  An anisotropic ensemble of bremsstrahlung-producing electrons will produce a radiation field that is not only polarized, but anisotropic. Measurements of the hard X-ray photon directivity can therefore provide a probe of the extent to which the accelerated electrons are beamed. 

One technique for studying hard X-ray directivity on a statistical basis is to look for center-to-limb variations.  Correlations between flare longitude and flare intensity or spectrum reflect the anisotropy of the X-ray emission and hence any directivity of the energetic electrons.  For example, if the radiation is preferably emitted in a direction parallel to the surface of the Sun, then a flare located near the limb will look brighter than the same flare near the disk center. Various statistical studies of X-ray flares at energies below 300 keV reported no significant center-to-limb variation of the observed intensity (e.g., \opencite{Datlowe74}; \opencite{Pizzichini74}; \opencite{Datlowe77}).  The conclusion from these early studies was that either the structure of the X-ray emitting region is variable from event to event or that the average motion of the energetic electrons is random.  However, a statistically significant center-to-limb variation in the spectral shape of the spectra of these events was found by \inlinecite{Roy75}, suggesting that perhaps some directivity may be present. 

The large sample of flares detected at energies greater than 300 keV by SMM GRS allowed, for the first time, a statistical search for directivity at higher energies.  Photons at these energies are well above the thermal emission from the flare. In addition, the greatly reduced Compton backscatter from the photosphere and the increased directionality of the bremsstrahlung cross-section make relatively strong radiation anisotropies a possibility at these energies. Analysis of high-energy data ($>300$ keV) from SMM GRS collected during cycle 21 provided the first clear evidence for directed emission, with a tendency for the high energy events to be located near the limb \cite{Vestrand87b,Bai88}. Using an independent set of flares detected by instruments on the Venera 13 and 14 spacecraft, \inlinecite{Bogovlov85} also reported evidence for anisotropies at hard X-ray energies. They also attributed this disk center-to-limb spectral variation to bremsstrahlung directivity. Observations from SMM GRS during cycle 22 provided further support for directivity \cite{Vestrand91}.  However, several high energy events were also observed near the disk center by a number of different experiments during cycle 22 (e.g., on GRANAT and CGRO; see Vilmer 1994 for a summary), perhaps suggesting a more complex pattern.  \inlinecite{Li95} used data from SMM to confirm the general results of \inlinecite{Vestrand87b} and concluded that there was evidence for increasing directivity with increasing energy. Quantifying the magnitude of the directivity from these statistical measurements is difficult.  For example, one needs to know the size-frequency distribution for flares as well as the form of the electron distribution to derive the predicted limb fraction (e.g. \opencite{Petrosian85}).  Furthermore, the results only represent an average for the flare sample.  Different flaring regions are not likely to have identical geometry.  Nor are individual flares likely to have time-independent electron distributions. 

Another method for studying the directivity in individual flares is the stereoscopic technique first employed by \inlinecite{Catalano73} at soft X-ray energies.  This method compares simultaneous observations made on two spacecraft that view the flare from different directions. \inlinecite{Kane88} combined observations from spec-trometers aboard the PVO (Pioneer Venus Orbiter) satellite and the ISEE-3 (Third International Sun-Earth Explorer) satellite to produce stereoscopic flare observations of 39 flares in the energy range from 100 keV to 1 MeV.  While the range of flux ratios measured by \inlinecite{Kane88} is consistent with the results of statistical studies \cite{Vestrand87a}, the deviations of the ratio from unity show no clear correlation with increasing difference in viewing angles.  \inlinecite{Li94} used sterescopic observations of 28 flares from the HXRBS and GRS experiments on SMM and the SIGNE experiments on Venera 13 and Venera 14 and concluded that directivity could only be marginally present in the 100--500 keV energy range.  Most recently, \inlinecite{Kane98} reported on a study of 13 flares in the 20--125 keV enery range using stereoscopic observations with X-ray instruments on both Ulysses and Yohkoh. They concluded that there was no clear evidence for directivity in the 20--125 keV energy range, although there were large calibration uncertainties present in the analysis. Stereoscopic observations require good calibration of individual instruments so that differences in the absolute fluxes observed by different instruments can be correctly interpreted. 

\subsection{Hard X-Ray Polarization Measurements}
The difficulties of statistical and stereoscopic observations for measuring hard X-ray directivity suggest the need for a technique that can measure time-dependent anisotropies for individual flares using a single instrument.  Polarization is a diagnostic that can meet these requirements.

Many models of nonthermal (e.g., thick target) hard X-ray production predict a clear and significant polarization signal, with polarization levels $>10$\% \cite{Brown72,Langer77,Bai78,Emslie80b,Leach83,Zharkova95,Charikov96}.  The precise level of polarization depends on both energy and viewing angle. Some fraction of the observed 20--100 keV hard X-ray flux will be flux that is backscattered from the solar photosphere, the precise magnitude of which will depend, in part, on the polarization of the primary flux.  The reflected component will, in turn, influence the degree of polarization of the observed flux, since even if the electrons are accelerated isotropically, backscattering will introduce polarization fractions of a few percent at energies below 100 keV (e.g., \opencite{Langer77}; \opencite{Bai78}).  Even thermal models of the hard X-ray source predict a finite polarization of order a few percent, due to the anisotropy in the electron distribution function caused by a thermal conductive flux out of the emitting region into the cooler surroundings \cite{Emslie80a}. The thermal component, with its rather low polarization, tends to dominate the emission from all flares at energies below about 25 keV.  At these energies, it therefore becomes difficult to distinguish the non-thermal component, with its intrinsic directivity signature, from the thermal component.  This has led to the argument that polarization measurements can best be performed at higher energies \cite{Chanan88}.

These predictions, while clearly testable, could be criticized on the grounds that the modeling assumptions they contain may be oversimplistic.  For example, each model to date assumes a single, simple magnetic field structure about which the emitting electrons spiral.  It could be argued that any real flare, particularly one sufficiently large to produce a signal of sufficient strength to enable a polarization measurement, will in all probability contain a mix of structures that would average out any polarization signal present.  However, hard X-ray imaging observations in the impulsive phase generally show a fairly simple geometry, consisting of two footpoint sources and perhaps a loop-top source (e.g., \opencite{Sakao92}; \opencite{Masuda95}). These observations suggest that simple magnetic structures are responsible for the energetic emissions and give support to the possibility that a statistically significant polarization signal could be produced in a large event.

The history of observations of hard X-ray polarization from solar flares is a fascinating subject in its own. The first measurements of X-ray polarization from solar flares (at energies of $\sim15$ keV) were made by Soviet experimenters using polarimeters aboard the Intercosmos satellites.  In their initial study, \inlinecite{Tindo70} reported an average polarization for three small X-ray flares of P = $40\% (\pm20\%)$. This study was followed by an analysis of three flares in October and November of 1970 \cite{Tindo72a,Tindo72b} that showed polarizations of approximately 20\% during the hard impulsive phase.  These reports were met with considerable skepticism, on the grounds that they did not adequately allow for detector cross-calibration issues and limited photon statistics \cite{Brown74}. Subsequent observations with an instrument on the OSO-7 satellite seemed to confirm the existence and magnitudes of the polarizations ($\sim~10\%$), but these data were compromised by in-flight gain shifts \cite{Nakada74}.  In a later study using a polarimeter on Intercosmos 11, \inlinecite{Tindo76} measured polarizations of only a few percent at $\sim15$ keV for two flares in July 1974. This small but finite polarization is consistent with the predictions for purely thermal emission that contains an admixture of polarized backscattered radiation \cite{Bai78}. A small polarimeter was flown on an early shuttle flight (STS-3) and made measurements of eight C- and M-class flares in the 5--20 keV energy range.  Upper limits in the range of 2.5\% to 12.7\% were measured, although contamination of the Li scattering material invalidated the pre-flight calibration \cite{Tramiel84}.    

The polarimetry capability of RHESSI will permit high quality observations of large flares. The RHESSI energy range for measuring polarization (extending to $\sim100$ keV) may be sufficiently high to reduce some of the thermal contamination that has plagued many previous measurements.  In addition, some fraction of the hard X-ray flux that will be observed by RHESSI will not be primary emission from the flare source, but rather will be flux that is backscattered from the solar photosphere.  These photospherically backscattered photons should form an ``albedo patch'' \cite{Brown75}, which may be directly imaged by RHESSI as a constraint on the contribution of such backscattered photons to the primary signal.  Clearly, the capability of RHESSI to simultaneuously image the hard X-ray emission represents a major advantage over previous efforts to measure hard X-ray polarization.

\section{Compton Polarimetry}

The basic physical process used to measure polarization in the 20--100 keV energy range is Compton scattering, the cross-section for which is given by the Klein-Nishina formula, 

	\begin{equation}
	\label{eq:1}
d\sigma = {r_{0}^{2} \over 2} d\Omega \left({\nu' \over \nu_o}\right)^2 
\left({\nu_o \over \nu'} + {\nu' \over \nu_o} - 2 \sin^2\theta \cos^2\eta \right)
	\end{equation}

\noindent where,

	\begin{equation}
	\label{eq:2}
{\nu' \over \nu_o} = {1 \over 1 + \left(h\nu_o \over mc^2 \right) \left(1 - \cos\theta \right)}
	\end{equation}

\noindent Here $\nu_o$ is the frequency of the incident photon,   $\nu'$ is the frequency of the scattered photon, $\theta$ is the scattering angle of the scattered photon measured from the direction of the incident photon, and $\eta$ is the azimuthal angle of the scattered photon measured from the plane containing the electric vector of the incident photon.  For a given value of $\theta$, the scattering cross section for polarized radiation reaches a minimum at $\eta = 0^{\circ}$ and a maximum at $\eta = 90^{\circ}$.  In other words, photons tend to be scattered at a right angle with respect to the incident electric field vector. In the case of an unpolarized beam of incident photons, there will be no net positive electric field vector and therefore no preferred azimuthal scattering angle ($\eta$); the distribution of scattered photons with respect to $\eta$ will be uniform.  However, in the polarized case, the incident photons will exhibit a net positive electric field vector and the distribution in $\eta$ will be asymmetric.  The asymmetry will be most pronounced for scatter angles near $\theta = 90^{\circ}$.

In general, a Compton scatter polarimeter consists of two detectors to determine the energies of both the scattered photon and the scattered electron (e.g., \opencite{Lei97}; \opencite{McConnell02}). One detector, the scattering detector, provides the medium for the Compton interaction to take place. This detector must be designed to maximize the probability of there being a single Compton interaction with a subsequent escape of the scattered photon.  This requires a low-Z material in order to minimize photoelectric interactions.  The primary purpose of the second detector, the calorimeter, is to absorb the full energy of the scattered photon.

\begin{figure}
\centering
\includegraphics[height=2.5in]{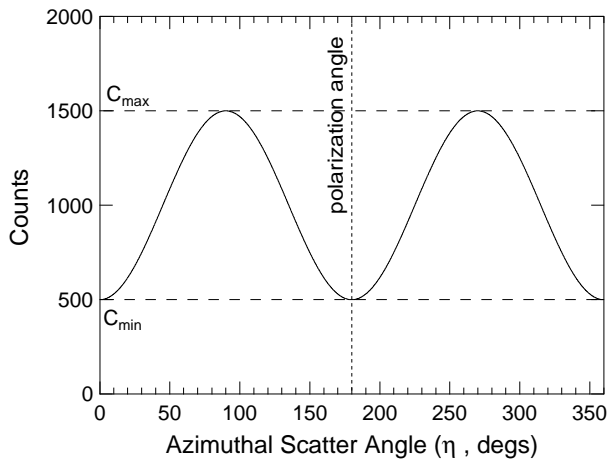}
\caption{An idealized modulation pattern produced by Compton scattering of polarized radiation.  The minimum of the modulation pattern defines the plane of polarization of the incident flux. }
\label{fig:ModPattern}
\end{figure}

The relative placement of the two detectors defines the scattering geometry. For incident photon energies below 100 keV, the azimuthal modulation of the scattered photons is maximized if the two detectors are placed at a right angle relative to the incident photon beam ($\theta = 90^{\circ}$). The two detectors must also be arranged relatively close to each other so that there is a finite solid angle for scattering to achieve the required detection efficiency. At the same time, a larger separation between the detectors provides more precise scattering-geometry information. The accuracy with which the scattering geometry can be measured determines the ability to define the modulation pattern (Figure~\ref{fig:ModPattern}) and therefore has a direct impact on the polarization sensitivity (see below). 

In principle, the scattering detector need not be active.  At soft X-ray energies ($<10$ keV), for example, the scattering process results in a very small energy loss that can be difficult to detect.   For polarimeters designed to measure soft X-rays, the scattering element is usually passive, its only requirement being that it be designed to maximize the probability of a single photon scatter.  At higher energies, an active element is certainly preferred as a means to minimize background, but it is certainly not required.

The ultimate goal of a Compton scatter polarimeter is to measure the azimuthal modulation pattern of the scattered photons.  From Equation (1), we see that the azimuthal modulation follows a $\cos 2\eta$ distribution.  More specifically, we can write the integrated azimuthal distribution of the scattered photons as,

	\begin{equation}
	\label{eq:3}
C(\eta) = A \cos ( 2 (\eta - \phi + {\pi \over 2})) + B 
	\end{equation}

\noindent where $\phi$ is the polarization angle of the incident photons; $A$ and $B$ are constants used to fit the modulation pattern (c.f., Figure~\ref{fig:ModPattern}). In practice, a measured distribution must also be corrected for geometrical effects based on the corresponding distribution for an unpolarized beam \cite{Lei97}.  The quality of the polarization signal is quantified by the polarization modulation factor \cite{Novick75,Lei97}.  For a given energy and incidence angle for an incoming photon beam, this can be expressed as,

	\begin{equation}
	\label{eq:4}
\mu_p = { C_{p,max} - C_{p,min} \over C_{p,max} + C_{p,min}}  = {A_p \over B_p}
	\end{equation}

\noindent where $C_{p,max}$ and $C_{p,min}$ refer to the maximum and minimum number of counts registered in the polarimeter, respectively, with respect to $\eta$; $A_p$ and $B_p$ refer to the corresponding parameters in Equation (3). In this case the $p$ subscript denotes that this refers to the measurement of a beam with unknown polarization. In order to determine the polarization of the measured beam, we need first to know how the polarimeter would respond to a similar beam, but with 100\% polarization.  This can be done using Monte Carlo simulations. We then define a corresponding modulation factor for an incident beam that is 100\% polarized,
	
		\begin{equation}
	\label{eq:5}
\mu_{100} = { C_{100,max} - C_{100,min} \over C_{100,max} + C_{100,min}}  = {A_{100} \over B_{100}}
	\end{equation}

Then, following \inlinecite{Novick75}, we can then use this result, in conjunction with the observed modulation factor ($\mu_p$), to determine the level of polarization in a measured beam,
 
		\begin{equation}
	\label{eq:6}
P = { \mu_p \over \mu_{100} }
	\end{equation}

\noindent where $P$ is the measured polarization. The minimum detectable polarization (MDP) can be expressed as \cite{Novick75},

		\begin{equation}
	\label{eq:7}
MDP(\%) = { n_{\sigma} \over \mu_{100} R_{src} } \sqrt{2 (R_{src} + R_{bgd}) \over T}
	\end{equation}

\noindent where $n_{\sigma}$ is the significance level (number of sigma), $R_{src}$ is the total source counting rate, $R_{bgd}$ is the total background counting rate and $T$ is the total observation time.

\section{The RHESSI Spectrometer}

\begin{figure}
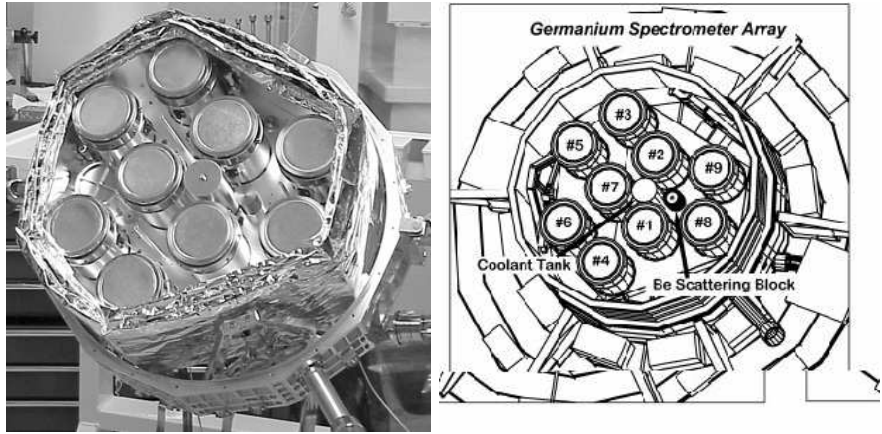

\centering
\includegraphics[height=2.25in]{FIGURE2a.EPSF}
\includegraphics[height=2.25in]{FIGURE2b.EPSF}
\caption{The RHESSI spectrometer array, as seen in a pre-flight photo (left) and as depicted by the GEANT3 mass model (right).}
\label{fig:spectrometer} 
\end{figure}

\begin{figure}
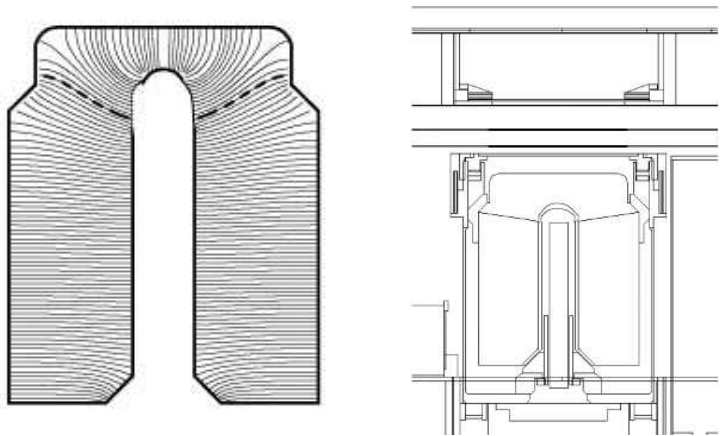

\centering
\includegraphics[height=2.25in]{FIGURE3a.EPSF}
\hspace{0.25in}
\includegraphics[height=2.25in]{FIGURE3b.EPSF}
\caption{Cross section of a RHESSI detector.  Left: A detector profile with electric field lines, the dashed field line indicating the separation between the front and back segments.  Right: A detector in the cryostat, showing Ta/Sn/Fe/Al shielding around the side of the front segment and above the shoulder of the rear segment.}
\label{fig:detector}
\end{figure}

The principal scientific objectives of RHESSI are being achieved with high resolution imaging spectroscopy observations from soft X-rays to gamma-rays \cite{Lin03}. The imaging system is made up of nine Rotating Modulation Collimators (RMCs), each consisting of a pair of widely separated grids mounted on the rotating spacecraft. The spectrometer (Figure~\ref{fig:spectrometer}) consists of nine segmented Germanium detectors, one behind each RMC, to detect photons from 3 keV to 17 MeV \cite{Smith03}. These detectors represent the largest, readily available, hyperpure (n-type) coaxial Ge material ($\sim7.1$ cm diam $\times 8.5$ cm long). The detectors are cooled to $\sim75^{\circ}$ K by a space-qualified long-life mechanical cryocooler.  Each detector (Figure~\ref{fig:detector}) is made from a single germanium crystal, which is electrically divided into independent front and rear segments to provide an optimum response for low and high energy photons, respectively.  The inner electrode is segmented into two contacts that collect charge from two electrically independent detector segments, defined by the electric field pattern. This provides the equivalent of a $\sim1$ cm thick planar Ge detector (the front segment) in front of a thick $\sim7$ cm coaxial Ge detector (the rear segment). 

The front segment thickness is chosen to stop photons incident from the front (solar-facing side) of the instrument up to $\sim100$ keV, where photoelectric absorption dominates, while minimizing the active volume for background. Front-incident photons that Compton-scatter, and background photons or particles entering from the rear, are rejected by anticoincidence with the rear segment.  A passive, graded-Z ring around the front segment (a laminate of Ta/Sn/Fe) absorbs hard X-rays incident from the side, to provide the low background of a phoswich-type scintillation detector.

Solar photons with energies from $\sim100$ keV to $\sim17$ MeV, including most nuclear gamma-ray lines, stop primarily in the thick rear segment alone. A smaller fraction of these high energy photons will stop in the front segment after first Compton scattering in the rear segment, thus depositing energy in both the front and rear segments.  Alternatively, some of these photons may deposit energy in two or more Ge detectors.  The intense 3--100 keV X-ray fluxes that usually accompany large $\gamma$-ray line flares are absorbed by the front segment, so the rear segment will always count at moderate rates. Photons with energy above 20 keV from non-solar sources can penetrate the thin aluminum cryostat wall from the side and also be detected by the Ge detector rear segments. 

RHESSI is a spinning spacecraft, with a spin rate of $\sim15$ rpm. The high angular resolution imaging capability of RHESSI imposes severe requirements on the knowledge of the instrument orientation direction at any given time.  Two spacecraft systems provide the necessary aspect solution. A solar aspect system provides knowledge of Sun center in pitch and yaw to 1.5 arcsec. A star scanner is used to sample the roll orientation at least once per rotation. Interpolation between star measurements allows the roll orientation to be determined with an accuracy of $\sim2.7$ arcmin. The energy and arrival time of every photon, together with spacecraft orientation data, are recorded in the spacecraft's on-board 4-Gbyte solid-state memory (sized to hold all the data from the largest flare) and automatically telemetered to the ground within 48 hours. These data can then be used to generate X-ray/$\gamma$-ray images with an angular resolution of $\sim2$ arcseconds and a FoV of $\sim1^{\circ}$.

\section{RHESSI as a Polarimeter}

It was realized during the RHESSI Phase-A study that the essential ingredients for measuring the polarization of solar flare hard X-rays, namely, an array of detectors in a rotating spacecraft, were already present in the RHESSI spectrometer.  All that was needed was the addition of a strategically placed cylinder of beryllium in the cryostat to Compton scatter the hard X-rays (20--100 keV) into the rear segments of the adjacent Ge detectors.  For the purpose of making polarization measurements, the important feature of the spectrometer design is that photons below 100 keV that are detected in the rear segments are either non-solar background or solar photons scattered into the rear segments (e.g., from the Be scattering block).  The distribution of events scattered into the adjacent rear segments (with respect to some fixed reference frame)  provides a polarization signature, since the direction of the scattering depends on the orientation of the electric field vectors, or plane of polarization, of the incoming photons.  The spacecraft rotation reduces the impact of systematic differences in detection efficiency amongst the different Ge detectors and increases the sampling frequency with respect to the azimuthal scatter angle ($\eta$ in Equation \ref{eq:1}).

The beryllium scattering block (3 cm in diameter by 3.5 cm long) was placed within the spectrometer cryostat, near the center of the Ge detector array (Figure~\ref{fig:spectrometer}).  Directly in front of the cryostat is a graded-Z shield (a laminate of Ta/Sn/Fe) that is designed to absorb a large fraction of the flux below 100 keV, flux which tends to dominate the flare event. Openings in this shield provide an unattenuated path for these low energy photons from the Sun to reach the front surface of the cryostat directly in front of each Ge detector and directly in front of the Be scattering block.  The cryostat aluminum is thinned in front of the Be scatterer (to get 20 keV threshold), but in front of the Ge detectors, it is cut out entirely and replaced with a thin Be window (to get 3 keV threshold).  In addition, the trays which hold the RMC grids each have a hole directly in front of the Be scattering block.  The total mass directly in front of the Be block amounts to $\sim1$ mm of aluminum in the cryostat plus some aluminized mylar thermal blankets, for a total mass of $\lesssim2$ mm aluminum equivalent.

The sensitive energy range for polarimetry with RHESSI is about 20--100 keV. The graded-Z shield in front of the cryostat insures that solar flare photons of this energy which enter the cryostat will interact either in the front surface of a Ge detector or in the Be block.  The segmented design of each Ge detector further insures that there are no direct solar flare photons in the rear segments within the 20--100 keV energy range.  Flare photons within this energy range can only reach the rear segments by first scattering within the Be block.  There is, however, a significant background from higher energy flare photons which scatter into the rear segments from other parts of the spacecraft and from flare photons which reflect off the Earth's atmosphere (the Earth albedo flux).  Photons that scatter into the Ge detectors from other parts of the spacecraft will generally not vary with spin angle.  This unmodulated component will interfere with the polarization measurement by effectively reducing the modulation factor.    

The albedo flux poses a potentially significant problem in that the level of albedo flux reaching a given Ge detector can vary significantly with spin phase (depending on the orientation of the spacecraft with respect to the local zenith). The magnitude of the albedo flux can be quite large, as much as 40--50\% of the direct solar flare flux at energies below 100 keV. Fortunately, the albedo flux has only 1 maximum per spin period instead of the two maxima exhibited by the polarization signal and can therefore be recognized using an appropriate analysis.  

The impact of the albedo flux may be somewhat further complicated by the fact that the precise distribution of the albedo flux across the visible disk of the Earth can also depend upon the polarization parameters of the direct flux from the flare. \citeauthor{McConnell96a}(\citeyear{McConnell96a,McConnell96b}) described this effect and how it might be used to measure the (energy-averaged) polarization parameters of hard X-rays from solar flares or $\gamma$-ray bursts using the BATSE detectors on CGRO.  While we expect that this effect will be quite difficult to observe with RHESSI, and hence unimportant in the polarization analysis, it is nontheless desirable to check this assumption with appropriate simulations.  If, by some chance, this effect could be measured by RHESSI, it would provide a useful complement to the direct polarization measurement.  

Finally, the variations in count rate within the detectors which measure the Be-scattered flux must be separated from intrinsic variations in the flare, but this can be done by examining the summed rate of the Ge detectors.  Any variations in count rate that may be related to the source modulation by the grids happen much faster than a spin period, and should average out for the polarization measurements.

\section{Simulated Polarization Response}

To simulate the polarimetric response of RHESSI, we have used a modified version of the GEANT3 code that includes the effects of polarization in Compton scattering and tracks the polarization of the incident photon. We make use of a detailed mass model developed by the RHESSI team.  This is a comprehensive mass model that includes not only the spectrometer, but also the complete telescope assembly, supporting electronics, spacecraft support structure and solar panels. 

In principle, each of the nine Ge detectors provides a polarization signature.  However, the quality of the signature varies, depending on the distance of the detector from the Be and the extent to which that detector is blocked from view with respect to the Be.  A particular detector may be blocked from view by another detector and/or by a cryogenic coolant tank (used only for ground operations and empty in flight).  Furthermore, the Ge detectors that are furthest from the Be do not provide a polarization signature with sufficient signal-to-noise to be useful in polarization studies.  The Ge detectors that are most useful in polarization studies are detectors 1, 2, 8 and 9 (c.f., Figure~\ref{fig:spectrometer}).  Unfortunately, detector 2 is currently not operating in segmentation mode and is therefore not usable for polarization studies.  The simulations presented below assume that only detectors 1, 8 and 9 are used.  

The initial simulations involved directing a narrow mono-energetic beam directly at the Be block.  In this case, the beam area corresponded to the area of the Be block (3 cm diameter).  These simulations allowed us to study the intrinsic polarimetric capabilities of RHESSI.  A second set of simulations used a much broader beam (60 cm diameter) that covered the full area of the telescope tube assembly.  These more realistic simulations allowed us to study the effects of photon scattering within the telescope tube and the surrounding spacecraft materials.  In this case, photon scattering refers to the solar photons which interact in rear Ge segments after scattering from some spacecraft component other than the Be.  These photons carry no polarization signature and serve only to increase the level of background.  

\begin{figure}
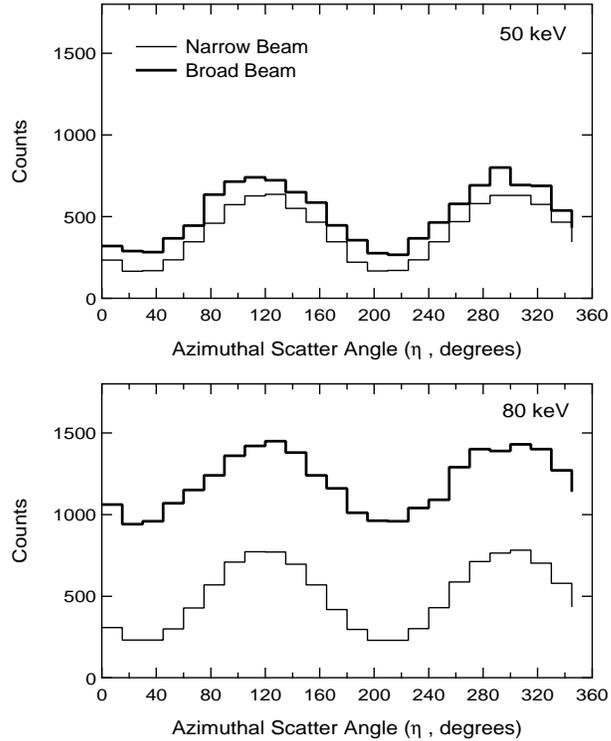

\centering
\includegraphics[width=3.25in]{FIGURE4a.EPSF}
\hspace{0.02\linewidth}
\includegraphics[width=3.25in]{FIGURE4b.EPSF}
\caption{The modulation patterns produced from both narrow and broad beam simulations for 50 keV (top) and 80 keV (bottom).  Although the total number of events increase dramatically at high energies using the more realistic broad beam simulation, most of these are unmodulated.}
\label{fig:narrow_vs_broad}
\end{figure}

\begin{figure}
\centering
\includegraphics[width=3.25in]{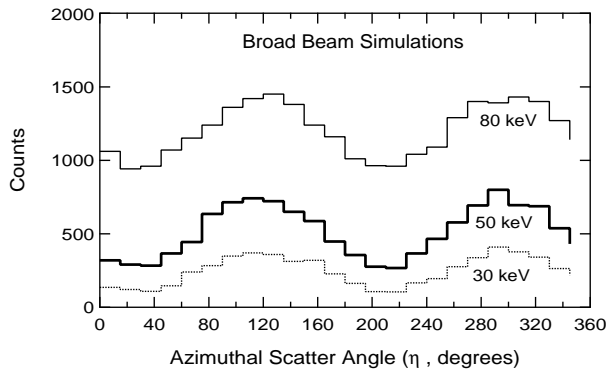}
\caption{The modulation patterns derived from the more realistic broad beam simulations for three different energies (30, 50 and 80 keV).  As energy increases, the total number of detected events increases, but the modulation factor decreases as a result of scattered (unmodulated) events.}
\label{fig:ModVsEnergy}
\end{figure}

Figure~\ref{fig:narrow_vs_broad} shows the simulated modulation patterns (number of counts versus azimuthal scatter angle) as derived from both the narrow and broad beam simulations. Only those events showing energy deposits above the 20 keV threshold in a single rear Ge segment (in detector 1, 8 or 9) are included.  At lower energies (50 keV), the modulation patterns are very similar.  But at higher energies (80 keV), the broad beam simulation shows that many more photons are detected.  Unfortunately, most of these photons are randomly scattered into the Ge detectors from material other than the Be.  These additional events therefore carry no useful polarization signature.   Figure~\ref{fig:ModVsEnergy} shows the modulation patterns for three different energies (30, 50 and 80 keV), as derived from the broad beam simulations.

We characterize the polarization response of RHESSI using two parameters: 1) the effective area ($A_{eff}$), which represents the effective area for events satisfying the necessary criteria (single energy deposit in rear segments of the selected Ge detectors);  and 2) the polarization modulation factor ($\mu_{100}$), which can be thought of as a measure of the quality of the polarization signature (Equation~\ref{eq:6}).  These two quantities can be used to define a third parameter, the figure-of-merit, that is useful in characterizing the polarization response,

	\begin{equation}
	\label{eq:9}
FoM = \mu_{100} \sqrt{A_{eff}} 
	\end{equation}

\begin{figure}
\centering
\includegraphics[width=2.75in]{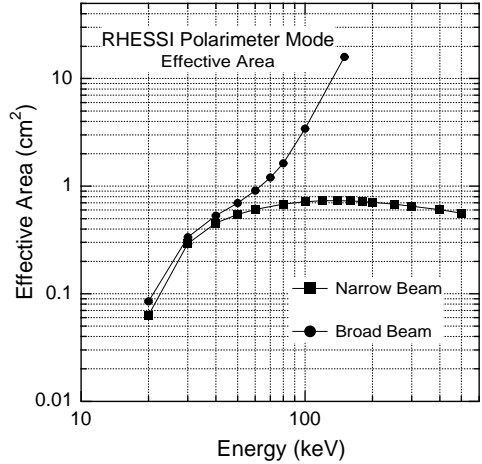}
\caption{Effective area of the polarimetry mode as a function of energy for both the narrow beam and broad beam simulations.  For the broad beam simulations, the effective area increases dramatically at higher energies, as a result of photons scattering into the Ge.}
\label{fig:EffArea_vs_Energy}
\end{figure}

\begin{figure}
\centering
\includegraphics[width=2.75in]{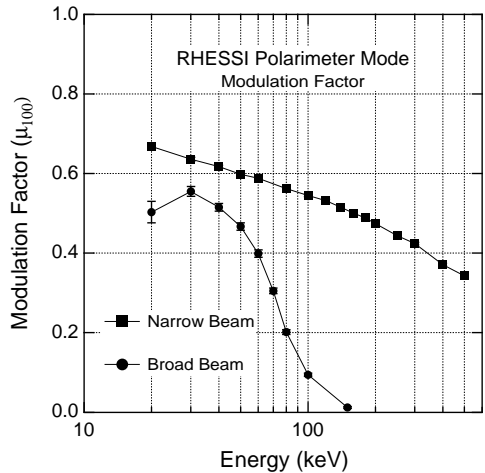}
\caption{Modulation factor of the polarimetry mode as a function of energy for both the narrow beam and broad beam simulations. For the broad beam simulations, the modulation factor decreases dramatically at higher energies, as a result of unmodulated photons scattering into the Ge.}
\label{fig:ModFct_vs_Energy}
\end{figure}

\begin{figure}
\centering
\includegraphics[width=2.75in]{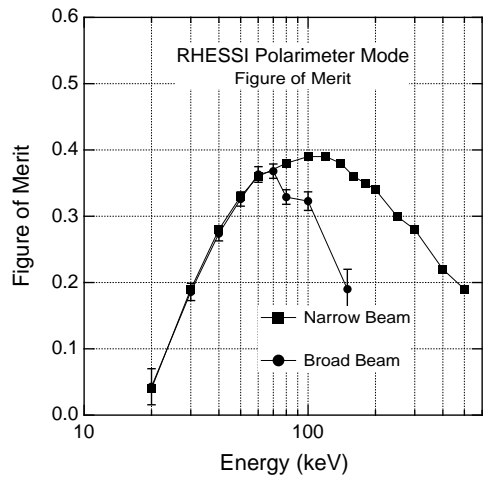}
\caption{Figure of merit of the polarimetry mode as a function of energy for both the narrow beam and broad beam simulations.}
\label{fig:FoM_vs_Energy}
\end{figure}

\noindent  As defined here, this figure-of-merit does not incorporate the effects of detector background (c.f., Figure~\ref{fig:bgd}). The characterization of the simulated response is shown in Figures~\ref{fig:EffArea_vs_Energy}--\ref{fig:FoM_vs_Energy}.  As before, we include only those events that deposit energy above threshold ($>20$ keV) in a single rear segment of detector 1, 8 or 9.  Of particular importance here is the significant impact of scattered photons, which is most easily seen in terms of the effective area (Figure~\ref{fig:EffArea_vs_Energy}).  The effective area is larger with the broad beam simulations as a result of the larger number of valid events due to photon scattering.  Above about 50 keV, scattered photons become increasingly important.  At energies near 100 keV and above, scattering completely dominates the response.  The effects of scattering are also seen in the plot of modulation factor versus energy (Figure~\ref{fig:ModFct_vs_Energy}).  Since the scattered component carries with it no polarization signature, the modulation factor decreases at energies above $\sim30$ keV, as scattering becomes more and more important.  Figure~\ref{fig:FoM_vs_Energy} shows that, in a source-dominated measurement, the efficacy of the RHESSI polarimeter mode  peaks near 60 keV.  

These results indicate that RHESSI is most effective as a polarimeter at energies below $\sim100$ keV.  One can also see that having an active scattering element, which would have allowed for rejection of scattered photons by selecting on only those events scattered in the Be, would have provided a very significant improvement in the polarimetric capability of RHESSI.   Although multiple scatter events between Ge detectors could be used for polarization measurements, the high-Z nature of the Ge will limit the sensitivity of this approach.  Nonetheless, we plan investigate this mode with future simulations.

\section{Polarization Sensitivity}

\begin{figure}
\centering
\includegraphics[width=3.6in]{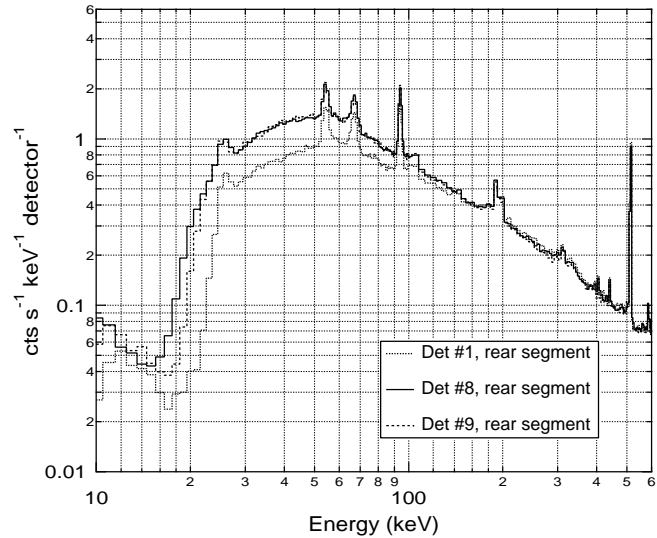}
\caption{Measured background spectra for RHESSI detectors 1, 8 and 9. This spectrum represents an average over about one orbit, excluding periods near the SAA and particle precipitation events. }
\label{fig:bgd}
\end{figure}

 The measured RHESSI background (Figure~\ref{fig:bgd}), along with the simulated polarimetric characteristics, can be used to estimate the polarization sensitivity for a typical solar flare.  Unfortunately, it is difficult to define a ``typical'' solar flare to use as a baseline for estimating polarization sensitivities.  Here we have used an X2-class flare having a spectrum of the form \cite{Chanan88},

	\begin{equation}
	\label{eq:10}
{dF \over dE} = 3.6 \times 10^7 E_{keV}^{-4.3} \ cm^{-2} s^{-1} keV^{-1} 
	\end{equation}

The X-ray classification depends only on the {\em peak} X-ray flux.  Therefore, the polarization sensitivity for a given class flare will depend on the duration of the event. We have used durations ranging from 20 seconds up to 500 seconds, with an average spectrum corresponding to that given above.  The results of these estimates are shown in Table~\ref{tab:MDP}.  Results for both an X2 class flare and an X10 class flare are shown, with the X10 spectrum assumed to have the same form as that above, but with a factor of 5 increase  in intensity.  The polarization sensitivity is given as the minimum detectable polarization (MDP; c.f., Equation~\ref{eq:7}).  The values represent a $3\sigma$ sensitivity in terms of percent polarization.

These estimates indicate that RHESSI has sufficient sensitivity to measure the polarization of X-class flares down to a level below 10\% and, in some cases, below 1\%.  This level of sensitivity will be useful in constraining various models that have been published in the literature.

Our sensitivity estimates have not yet considered the effects of the atmospheric albedo flux.  As much as 30--40\% of the solar flux gets Compton scattered off the atmosphere at these energies.  If the level of this component reaches a level comparable to the intrinsic detector background (as might be expected), the quoted sensitivities would be about 50\% worse than shown.  Even such a degraded sensitivity will still provide important scientific results.

\begin{table}
\caption{Minimum Detectable Polarization} 
\label{tab:MDP}
\begin{tabular}{cccc} 
\hline
\rule[-1ex]{0pt}{3.5ex} & \multicolumn{3}{c} {{\it Event Duration}} \\
\rule[-1ex]{0pt}{3.5ex} & 20 sec & 100 sec & 500 sec \\
\hline
\rule[-1ex]{0pt}{3.5ex}{\it X2 class flare} &  & &  \\
\rule[-1ex]{0pt}{3.5ex}20--40 keV & 11\% & 5\% &  2\% \\
\rule[-1ex]{0pt}{3.5ex}40--60 keV & 53\% & 24\% &  11\% \\
\rule[-1ex]{0pt}{3.5ex}60--80 keV & -- & -- &  46\% \\
\hline
\rule[-1ex]{0pt}{3.5ex} {\it X10 class flare} &  & &  \\
\rule[-1ex]{0pt}{3.5ex}20--40 keV & 5\% & 2\% &  1\% \\
\rule[-1ex]{0pt}{3.5ex}40--60 keV & 17\% & 7\% &  3\% \\
\rule[-1ex]{0pt}{3.5ex}60--80 keV & 61\% & 27\% &  12\% \\
\hline
\end{tabular}
\end{table}

\section{Prospects for Solar Flare Polarization Studies}

The RHESSI spacecraft was successfully launched on February 5, 2002.  To date, there have been several M- and X-class flares that are good candidates for polarization studies.  We are currently working to develop the detailed procedures for analyzing these data.  These efforts are concentrating first on data from the the X4.8 event of 23-July-2002, which remained at the X-class level of emission for at least 900 seconds, and the M4.0 event of 17-March-2002, which occurred near local noon where modulation of the albedo flux should be minimal.  Given the quality of these data, along with the predicted sensitivity levels, we anticipate being able to achieve sensitivity levels below 10\% for some of these events.

\begin{acknowledgements}

This work has been supported by NASA grant NAG5-10203 at UNH and UAH, and by NASA contract NAS5-98033 at UC Berkeley.

\end{acknowledgements}

\theendnotes

\end{article}
\end{document}